\documentclass[a4paper,12pt]{article}
\usepackage{amsmath,amsfonts,amssymb,cite}

\begin{document}

\begin{center}
{\Large\bf Soft wall model with inverse exponential profile as a
model for the axial and pseudoscalar mesons}
\end{center}

\begin{center}
{\large S. S. Afonin\footnote{E-mail: afonin@hep.phys.spbu.ru.}}
\end{center}

\begin{center}
{\it V. A. Fock Department of Theoretical Physics, Saint-Petersburg
State University, 1 ul. Ulyanovskaya, 198504, Russia}
\end{center}

\begin{abstract}
According to the conjecture of the gauge/gravity duality a strongly
coupled 4D field theory may be described by a dual 5D theory which
is a free field theory in the first approximation. The usual
description of the axial-vector sector in the bottom-up AdS/QCD
models includes, however, the quartic interaction with a bulk scalar
field. The question appears whether it is possible to describe some
aspects of the chiral symmetry breaking in a way more consistent
with the original proposal, {\it i.e.} within a free 5D theory? We
suggest that a natural candidate for this purpose is the soft-wall
model with inverse exponential profile.
\end{abstract}

\section{Introduction}

The ideas of AdS/CFT correspondence from the string
theory~\cite{mald,witten,gub} have inspired the appearance of
holographic approach to QCD, in particular of the bottom-up
holographic models. These models try to describe the
non-perturbative dynamics of strong interactions in terms of a
putative semiclassical five-dimensional theory. The first versions
of bottom-up models describing the meson spectrum and the Chiral
Symmetry Breaking (CSB) --- the so-called hard wall
models~\cite{son1,pom}
--- had incorrect predictions for the behavior of spectrum of radial
meson excitations ($m_n\sim n$ instead of the string like behavior
$m_n^2\sim n$ expected in the phenomenology~\cite{phen}) and for the
non-perturbative corrections to the leading parton logarithm of
two-point correlation functions (exponential instead of expected
polynomial corrections). The introduction of the Soft Wall (SW)
holographic model~\cite{son2} has solved these problems.

The situation with a self-consistent description of the CSB is more
complicated. To reflect the Goldstone theorem and related
phenomenology like the mass splitting between the vector and
axial-vector states one introduces the interaction term(s) in a 5D
holographic action. The resulting models are at odds with the
initial assumption that non-perturbative dynamics of a strongly
coupled 4D gauge theory, in the first approximation, admits a dual
description in terms of a free 5D field theory. Let alone a sad fact
that following this way one abandons the hope to learn something new
about the strongly coupled QCD from the holographic approach.
Instead of this one attempts to find a 5D description of old known
physics that is not worse than the descriptions given by the old
known effective theories for QCD.

Within the usual effective models describing the strong interactions
at low energies, the dynamics responsible for the CSB is separated
from the dynamics responsible for the confinement. The latter
phenomenon is not necessary present in the effective quark models
(for instance, it is absent in the Nambu--Jona-Lasinio like models).
{\it A priori} it is not evident why the separation of the CSB from
the confinement can be done in the 5D dual theories. Since the
holographic models yield colorless mesons at any energies the
confinement is definitely present by construction. It is not
excluded, however, that a part of the CSB physics is encoded in the
choice of 5D background. If this is the case then by a successful
choice of 5D background we can catch some important aspects of the
CSB on the level of a free holographic theory.

In building the holographic models for QCD it should be always kept
in mind that the conjectural outcome of these models is given by the
correlators of currents in the large-$N_c$ limit. They encode
various dynamical information. The spectrum is a part of this
information. The equations of motion provide just the alternative
way to find the spectrum without calculating the two-point
correlators. The correlators represent thus the primary objects
which must be compared with their QCD counterparts. It is known that
the axial-vector two-point correlator should contain a massless (in
the chiral limit) pion pole. This is an important consequence of the
CSB. Basing on this observation we suggest that the free action of
the SW model with inverse dilaton is a natural candidate for the
description of the axial-vector channel, with the CSB taken into
account in the first approximation. In the present work, we
elaborate this point.

The paper is organized as follows. The basics of the SW models are
reminded in the next section, where we discuss the advantages and
disadvantages of these models. In Sect.~3, it is shown that the SW
model with inverse dilaton is a good first approximation for the
axial channel. The improvement of the SW model with inverse dilaton
is presented in Sect.~4. The Sect.~5 contains a brief discussion of
the scalar sector. Then, in Sect.~6, we make some remarks and
conclude in Sect.~7.

\section{The soft wall model}

In this section, we briefly summarize the main results of the
holographic SW model. The simplest action of the SW model describing
the vector mesons is given by
\begin{equation}
\label{1a}
S=-\frac{c^2}{4}\int d^4\!x\,dz\sqrt{g}\,e^{-az^2}F_{MN}F^{MN},
\end{equation}
where $F_{MN}=\partial_M V_N-\partial_N V_M$, $M=0,1,2,3,4$ (the
metric is mostly negative), and $c$ represents a normalization
constant for the field $V_M$. The action~\eqref{1a} is defined in
the AdS$_5$ space whose metric can be parametrized as follows,
\begin{equation}
\label{2a}
ds^2=\frac{R^2}{z^2}(dx_{\mu}dx^{\mu}-dz^2),\qquad z>0,
\end{equation}
here $R$ is the radius of the AdS$_5$ space and $z$ is the
holographic coordinate. On the boundary of the AdS$_5$ space,
$z=\epsilon\rightarrow0$, the field $V_M$ corresponds to the source
for a QCD operator interpolating the vector mesons,
$V_M(x,\epsilon)\leftrightarrow\bar{q}\gamma_{\mu}q$ or
$V_M(x,\epsilon)\leftrightarrow\bar{q}\gamma_{\mu}\gamma_5q$.
According to the prescriptions of the AdS/CFT
correspondence~\cite{witten,gub}, the masses of fields in a dual
theory defined on the AdS$_5$ space are
\begin{equation}
\label{3a}
m_5^2R^2=(\Delta-J)(\Delta+J-4),
\end{equation}
where $\Delta$ is the canonical dimension of the corresponding 4D
field theory operator and $J$ is the spin, $J=0,1$. We will set
$R=1$ in what follows. In the case under consideration, $\Delta=3$.
This canonical dimension for the vector current results in the zero
mass for the 5D vector field $V_M$.

Recently some questions appeared concerning the choice of the sign
of the exponential profile (called also dilaton profile) presenting
in the action of the SW models. In particular, it was proposed that
the inverse (with respect to the choice of Ref.~\cite{son2}) dilaton
profile provides nicer confinement properties~\cite{br3,zuo} and
better description of the CSB~\cite{zuo}. However, the SW model with
inverse dilaton leads to the appearance of massless pole in the
vector correlator which cannot be eliminated~\cite{son3}.

The action~\eqref{1a} is purely phenomenological. It is not known
which 5D dynamical model leads to the particular background
$e^{-az^2}$ as a solution of 5D Einstein equations. An
"intermediate" dynamical model that leads to such a background could
look as follows,
\begin{equation}
\label{3b} S=\int
d^4\!x\,dz\sqrt{g}(\partial_M\!\varphi\partial^M\!\varphi-m^2\varphi^2+e^{\varphi}\mathcal{L}),
\end{equation}
where the 5D space is the AdS one and $\mathcal{L}$ represents some
Lagrangian density. If the 5D mass squared $m^2$ of the "dilaton"
field $\varphi$ takes the minimal value permitted in the AdS space,
$m^2=-4$ (the Breitenlohner-Freedman stability
bound~\cite{freedman}), then the equation of motion
(see~\eqref{24a}) will have a solution $\varphi=-az^2$. The sign of
the constant $a$ is not fixed. In fact, it is difficult to give
dynamical arguments which would allow to fix this sign. On the
heuristic level, since $z$ is the inverse energy scale one may
imagine that the background $e^{-az^2}$ acts as a "projector": The
choice $a<0$ enhances the part of the action that is important in
the infrared limit. Thus, the scalar part should contain the
pseudoscalar mesons and the vector part should contain the
axial-vector field because the pions emerge from the divergence of
the axial-vector current. The same heuristic argument suggests the
choice $a>0$ for the description of the scalar and vector states.

The hypothesis of AdS/CFT correspondence yields a practical tool for
calculation of correlation functions in a strongly coupled 4D field
theory~\cite{witten,gub}. All such functions can be obtained from
the generating functional of the connected correlators
$W_{4D}[\varphi_0(x)]$ that depends on the sources $\varphi_0(x)$
for the 4D field theory operators. If the 5D dual theory exists, the
holographic correspondence postulates the identification
\begin{equation}
\label{4a}
W_{4D}[\varphi_0(x)]=S_{5D}[\varphi(x,\epsilon)].
\end{equation}
The 5D dual theory is assumed to be in the weakly coupled regime.
This implies two important consequences: As the first approximation,
we may restrict ourselves by the quadratic terms in $S_{5D}$ and use
the semiclassical analysis. Thus, roughly speaking, the recipe for
calculation of n-point correlators is as follows: Evaluate $S_{5D}$
on the solution of equation of motion and differentiate n times with
respect to boundary values of 5D fields.

Let us apply this recipe to the action~\eqref{1a}. This action is
gauge invariant so we may use the gauge freedom to fix the axial
gauge, $V_z=0$, which is the most convenient. In addition, we will
consider the transverse fields, $\partial_{\mu}V^{\mu}=0$. The
dependence on the usual 4D coordinates of field $V^{\mu}(x,z)$ can
be Fourier transformed to $V^{\mu}(q,z)$. If we let
$V^{\mu}(q,z)=v(q,z)V^{\mu}_0(q)$ and impose $v(q,\epsilon)=1$ then
$V^{\mu}_0(q)$ can be interpreted as the source for the operator of
vector current. The equation of motion for the scalar function
$v(q,z)$ follows from the action~\eqref{1a},
\begin{equation}
\label{5a}
\partial_z\left(\frac{e^{-az^2}}{z}\partial_z v\right) + \frac{e^{-az^2}}{z}q^2v=0.
\end{equation}
Evaluation of the action~\eqref{1a} on the solution of
Eq.~\eqref{5a} yields the boundary term
\begin{equation}
\label{6a}
S=\frac{c^2}{2}\int d^4\!xV^{\mu}_0V_{0\mu}\left.\frac{e^{-az^2}}{z}\,v\partial_z v\right|^{z=\infty}_{z=\epsilon}.
\end{equation}

The two-point correlator $\Pi_V(-q^2)$ of vector currents $J_{\mu}$ is defined as
\begin{equation}
\label{7a}
\int d^4x e^{iqx}\langle
J_{\mu}(x)J_{\nu}(0)\rangle=(q_{\mu}q_{\nu}-q^2g_{\mu\nu})\Pi_V(-q^2).
\end{equation}
It can be found by differentiating twice with respect to source
$V^{\mu}_0$ in~\eqref{6a} near the boundary $z=\epsilon$,
\begin{equation}
\label{8a}
\Pi_V(-q^2)=c^2\left.\frac{\partial_z v}{q^2z}\right|_{z=\epsilon}.
\end{equation}
Solution of Eq.~\eqref{5a} with the boundary conditions $v(q,\epsilon)=1$, $v(q,\infty)=0$ is
\begin{equation}
\label{9a}
v(q,z)=\Gamma\left(1-\frac{q^2}{4|a|}\right)e^{(a-|a|)z^2/2}U\left(\frac{-q^2}{4|a|},0;|a|z^2\right),
\end{equation}
where $U$ is the Tricomi confluent hypergeometric function.
Substitution of~\eqref{9a} in~\eqref{8a} results in
\begin{equation}
\label{10a}
\Pi_V(-q^2)=c^2\left[\frac{a-|a|}{q^2}-\frac12\psi\left(1-\frac{q^2}{4|a|}\right)\right]+\text{const}.
\end{equation}
The digamma function $\psi$ has the following representation
\begin{equation}
\label{11a}
\psi\left(1-\frac{q^2}{4|a|}\right)=\sum_{n=0}^{\infty}\frac{4|a|}{q^2-4|a|(n+1)}+\text{const}.
\end{equation}
Using~\eqref{11a} one arrives at the pole representation for the
vector correlator,
\begin{equation}
\label{12a}
\Pi_V(-q^2)=c^2\left[\frac{a-|a|}{q^2}+\sum_{n=0}^{\infty}\frac{2|a|}{4|a|(n+1)-q^2}\right]+\text{const}.
\end{equation}
The poles in the sum of expression~\eqref{12a} give the mass
spectrum of the model\footnote{The independence of spectrum of
massive states on the sign of $a$ is a peculiarity of the vector
channel. For instance, a straightforward calculation shows that the
spectrum of the scalar mesons described by the quadratic in fields
5D action does not have this property (see Eq.~\eqref{33a}).}.

As is seen from the expression~\eqref{12a}, the sign choice for $a$
is important. If the sign choice is negative, $a<0$, the vector
correlator contains the massless pole. The physical origin of this
pole has been nicely discussed in Ref.~\cite{son3}. Rephrasing the
essence in a short way, the vector physical modes are looked for in
the form
\begin{equation}
\label{13a}
V_{\mu}(q,z)=\varepsilon_{\mu}e^{iqx}v(z),
\end{equation}
where $\varepsilon_{\mu}$ denotes the polarization vector, $e^{iqx}$
is the 4D plane wave, and $v(z)$ represents a profile depending on
the holographic coordinate. The spectrum of massive modes
corresponds to normalizable solutions of Eq.~\eqref{5a} whose
eigenvalues yield $q^2_n=m_n^2$. These solutions satisfy the
boundary condition $v(\epsilon)=0$ and leave the action~\eqref{6a}
finite. However, there is a massless non-normalizable solution
$v(z)=e^{az^2}$ which also leaves the action~\eqref{6a} finite if
$a<0$.

Note that at large euclidean momentum $Q^2=-q^2$ the
correlator~\eqref{10a} has the following asymptotic expansion
\begin{equation}
\label{14a}
\Pi_V(Q^2)_{Q^2\rightarrow\infty}=\frac{c^2}{2}\left[\log\left(\frac{4|a|}{Q^2}\right)
-\frac{2a}{Q^2}+\frac{4a^2}{3Q^4}+\mathcal{O}\left(\frac{a^4}{Q^8}\right)\right].
\end{equation}
On the other hand, in QCD sum rules, the Operator Product
Expansion (OPE) for the same correlator reads~\cite{svz},
\begin{equation}
\label{15a}
\Pi_V(Q^2)_{\text{OPE}}=\frac{N_c}{24\pi^2}\log\left(\frac{\mu^2}{Q^2}\right)+
\frac{\alpha_s}{24\pi}\frac{\langle G^2\rangle}{Q^4}+
\xi\frac{\langle\bar{q}q\rangle^2}{Q^6}+
\mathcal{O}\left(\frac{\mu^8}{Q^8}\right),
\end{equation}
where $\langle G^2\rangle$ and $\langle\bar{q}q\rangle$ mean the
gluon and quark condensate, $\mu$ is a renormalization scale, and
the constant $\xi$ is different for the vector and axial-vector
case. Comparison of expressions~\eqref{14a} and~\eqref{15a} leads to
the following conclusions: a) the normalization factor $c$ for the
5D vector field is fixed,
\begin{equation}
\label{15b}
c^2=\frac{N_c}{12\pi^2};
\end{equation}
b) the sign of $\langle G^2\rangle$ is reproduced correctly; c) the
$\mathcal{O}(Q^{-6})$ contribution is absent as we expected since
the effect of quark condensate was not included into the model; d)
the presence of $\mathcal{O}(Q^{-2})$ contribution disagrees with
QCD as long as its numerator would correspond to a local
gauge-invariant condensate of dimension two which cannot be
constructed in QCD. The last point represents a serious drawback of
the SW models that seems to be ignored in the
literature\footnote{Except the Ref.~\cite{ijmpa26} where this
observation was the main motivation to modify the SW model.}.

\section{The model with inverse dilaton corresponds to the axial mesons}

The vector spectrum of the SW model is independent of the sign of
slope parameter $a$. However, the choice $a<0$ leads to the massless
state in the vector channel that contradicts QCD. For this reason
the model with $a<0$ was discarded in the original
paper~\cite{son2}, where the SW model was introduced. On the other
hand, it is known that the axial-vector channel does contain the
massless (in the chiral limit) pseudoscalar state due to the Partial
Conservation of Axial Current (PCAC) hypothesis. Consequently, the
model with inverse dilaton, $a<0$, may be interpreted as a natural
SW model for the axial-vector mesons. We wish to present some
additional arguments in favor of our conjecture.

First of all, the spectrum of the model behaves as $m^2\sim n+1$.
Theoretically, such a spectrum is typical for the axial-vector
mesons while the vector mesons should have $m^2\sim n+1/2$ with the
same slope. This pattern of spectrum holds in the generalized
Lovelace-Shapiro dual amplitude~\cite{avw} which was
phenomenologically the most successful among the dual amplitudes of
the Veneziano type. Also it appeared in the early versions of QCD
planar (= large-$N_c$) sum rules~\cite{kat1}. The comparison with
the latter method is crucial. Concerning the problems of meson
spectroscopy, the bottom-up holographic models may be interpreted as
an exact five-dimensional reformulation of QCD planar sum
rules~\cite{afoninI}. Choosing the negative sign for the parameter
$a$ we incorporate an important feature of the chiral symmetry
breaking in QCD --- the PCAC. As a result, the residue in the
massless pole must be interpreted as $f_{\pi}^2$, where
$f_{\pi}=92.4$~MeV is the pion decay constant. But we do not yet
incorporate another order parameter of CSB --- the quark condensate.
In the original SVZ sum rules~\cite{svz}, the latter had impact on
the masses of mesons on the level of several percent only. Thus,
except the pseudoscalar states, discussing the spectrum of mesons in
the large-$N_c$ limit of QCD, it seems to be a good first
approximation to neglect the quark condensate. Saturating the
two-point correlators by the spectrum linear in masses squared, the
planar QCD sum rules can be solved in this approximation, the result
for the vector ($V$) and axial-vector ($A$) correlators
is~\cite{sr6} (possible corrections to this picture are considered
in Ref.~\cite{we})
\begin{equation}
\label{16a}
\Pi_V(Q^2)=\sum_{n=0}^{\infty}\frac{2f_{\pi}^2}{Q^2+\Lambda(n+1/2)}+\text{const},
\end{equation}
\begin{equation}
\label{17a}
\Pi_A(Q^2)=\frac{f_{\pi}^2}{Q^2}+\sum_{n=0}^{\infty}\frac{2f_{\pi}^2}{Q^2+\Lambda(n+1)}+\text{const},
\end{equation}
where
\begin{equation}
\label{18a}
\Lambda=\frac{48\pi^2}{N_c}f_{\pi}^2.
\end{equation}
The spectrum has the pattern mentioned above and is parametrized
completely by the constant $f_{\pi}$. The result~\eqref{18a} follows
for our slope $4|a|$ from the comparison of residues of massive
states in~\eqref{12a} and~\eqref{17a} taking into account the
normalization factor~\eqref{15b}. However, the massless residue in
the correlator~\eqref{12a} for $a<0$ turns out to be twice the
massless residue in the axial correlator~\eqref{17a}.

Thus, interpreting the SW model with $a<0$ as a model for the
axial-vector mesons we encounter two problems: (1) the presence of
dimension-two condensate; (2) the value of pion residue is too
large. Both problems disappear if we reformulate the SW model as a
model without the dilaton profile. This model will be considered in
the next section.

From the phenomenological point of view, the spectrum of the SW
model,
\begin{equation}
\label{18b}
m_n^2=m_0^2(n+1),\qquad n=0,1,2,\dots.
\end{equation}
also corresponds to the axial-vector mesons rather than to the
vector states. For comparison of the spectrum~\eqref{18b} with the
experimental data it is convenient to display it in units of mass
squared of the ground state,
\begin{equation}
\label{18c}
m_n^2=m_0^2\{1,2,3,4,\dots\}.
\end{equation}
According to the Particle Data~\cite{pdg}, the spectrum of
$\rho$-mesons is (in MeV, experimental errors are neglected): $775$,
$1465$, $1720$, $1900^{[?]}$, $2000^{[??]}$, $2149^{[?]}$,
$2265^{[??]}$, where the sign $[?]$ marks the states which need
confirmation and the sign $[??]$ marks badly established resonances
which were usually seen by one collaboration only. Using the
form~\eqref{18c} this data can be displayed as follows,
\begin{equation}
\label{18e}
m_{\rho,n}^2=m_{\rho}^2\{1,3.6,4.9,6.0^{[?]},6.7^{[??]},7.7^{[?]},8.5^{[??]}\}.
\end{equation}
Comparison of~\eqref{18e} with~\eqref{18c} does not reveal
similarity.

Consider the spectrum of known axial-vector states $a_1$~\cite{pdg}:
$1230$, $1640^{[?]}$, $1930^{[??]}$, $2265^{[??]}$. In the form
of~\eqref{18c}, it reads
\begin{equation}
\label{18f}
m_{a_1,n}^2=m_{a_1}^2\{1,1.8^{[?]},2.5^{[??]},3.4^{[??]}\}.
\end{equation}
The spectrum~\eqref{18f} resembles~\eqref{18c}, at least
qualitatively.

A better quantitative agreement with the axial-vector spectrum is
achieved if one introduces the ultraviolet cutoff
$z=\Lambda_{\text{cut}}$ to the SW model. The introduction of
$\Lambda_{\text{cut}}$ looks reasonable as long as QCD is weakly
coupled in the ultraviolet domain, hence, its hypothetical
holographic dual theory should be strongly coupled in that domain
and the semiclassical treatment with the latter is not justified any
more. The vector correlator of the SW model with the ultraviolet
cutoff was calculated in Ref.~\cite{afonin},
\begin{equation}
\label{18g}
\Pi_V(-q^2)=c^2\frac{e^{-|a|\Lambda_{\text{cut}}^2}}{2}\frac{U(1-q^2/(4|a|),1,|a|\Lambda_{\text{cut}}^2)}
{U(-q^2/(4|a|),0,|a|\Lambda_{\text{cut}}^2)}.
\end{equation}
The poles of correlator~\eqref{18g} yield the mass spectrum which
becomes now nonlinear. In particular, the mass of the first excited
state gets closer to the mass of the ground one, $m_1^2/m_0^2<2$.
This corresponds indeed to the experimental axial-vector spectrum
and is contrary to the situation in the vector one. For instance,
the choice $|a|\Lambda_{\text{cut}}^2=1$ gives rise to the following
spectrum,
\begin{equation}
m_n^2=m_0^2\{1,1.8,2.6,3.3,4.1,\dots\},
\end{equation}
which agrees very well with the experimental data on the
$a_1$-mesons~\eqref{18f}.

\section{Improvement: The no-wall model}

In this Section, we indicate a possible way to solve the problems
(1) and (2) of the standard SW model. In fact, this way has been
already proposed in Ref.~\cite{ijmpa26}. We will repeat the main
steps and add some new points which were not mentioned in
Ref.~\cite{ijmpa26}.

Let us remove the dilaton profile in the action~\eqref{1a} with the
help of the transformation~\cite{ijmpa26}
\begin{equation}
\label{19a}
V_M=e^{az^2/2}\tilde{V}_M.
\end{equation}
The action becomes equivalent to (we will not write the normalization
constant)
\begin{equation}
\label{20a}
S=-\frac{1}{4}\int d^4\!x\,dz\sqrt{g}\left(\tilde{F}_{MN}\tilde{F}^{MN}+
\frac{a^2z^4}{2}\tilde{V}_{\mu}\tilde{V}^{\mu}\right)+
\frac{a}{2}\int d^4\!x\left.\tilde{V}_{\mu}^2\right|_{z=0}^{z=\infty}.
\end{equation}
The last surface term was omitted in Ref.~\cite{ijmpa26}. We keep it
for the reason that now will be clear. The holographic profile of
physical modes~\eqref{13a} has the following behavior
\begin{equation}
\label{21a}
v(z)\sim z^ke^{(a-|a|)z^2/2},\qquad k>0.
\end{equation}
The surface term in the action~\eqref{20a} disappears if $a<0$ and
is infinite if $a>0$. Hence, the reformulation of the SW model of
the kind~\eqref{20a} is possible only for the case $a<0$.

As was demonstrated in Ref.~\cite{ijmpa26}, the vector correlator of
the no-wall model is (we insert again the normalization factor)
\begin{equation}
\label{26a}
\Pi_V(-q^2)=c^2\left[-\frac{|a|}{q^2}+\sum_{n=0}^{\infty}\frac{2|a|}{4|a|(n+1)-q^2}\right]+\text{const},
\end{equation}
with the asymptotic expansion at large $Q^2=-q^2$
\begin{equation}
\label{27a}
\Pi_V(Q^2)_{Q^2\rightarrow\infty}=\frac{c^2}{2}\left[\log\left(\frac{4|a|}{Q^2}\right)
+\frac{4a^2}{3Q^4}+\mathcal{O}\left(\frac{a^4}{Q^8}\right)\right].
\end{equation}
Now the massless residue is the half of massive residue and the
$\mathcal{O}(Q^{-2})$ contribution in the expansion~\eqref{27a} is
absent. Thus, both our goals are achieved simultaneously.

The price to pay is the appearance of the $z$-dependent mass term in
the action~\eqref{20a}. This term breaks the original gauge
invariance. There is an easy trick that restores the gauge
invariance: the emerging $z$-dependent mass may be interpreted as an
effect of condensation of some bulk scalar field which is coupled to
the vector field via the covariant derivative (we omit tildes
henceforth),
\begin{equation}
\label{22a}
S=\int d^4\!x\,dz\sqrt{g}\left(|D_M\varphi|^2-m_{\varphi}^2\varphi^2
-\frac{1}{4}F_{MN}F^{MN}\right),
\end{equation}
where
\begin{equation}
\label{23a}
D_M=\partial_M-i\lambda V_M.
\end{equation}
The action~\eqref{22a} contains the quartic interaction which we
have tried to avoid. Nevertheless, let us see how this trick works.
The equation of motion for $\varphi$ in the absence of the vector
field $V_M$ is
\begin{equation}
\label{24a}
-\partial_z\left(\frac{\partial_z
\varphi}{z^3}\right)+\frac{m_{\varphi}^2\varphi}{z^5}=0.
\end{equation}
To give the desired condensate term it must have a solution
$\varphi_0\sim z^2$. This takes place if $m_{\varphi}^2=-4$. The
naive application of relation~\eqref{3a} results in the conclusion
that the field $\varphi$ corresponds to an operator of canonical
dimension $\Delta=2$. This conclusion is debatable. According to an
AdS/CFT prescription~\cite{kleb}, the solution of classical equation
of motion for a scalar field $\Phi$ corresponding to an operator $O$
has the following form near the 4D boundary $z\rightarrow0$,
\begin{equation}
\label{25a}
\Phi(x,z)_{z\rightarrow0}=z^{4-\Delta}\Phi_0(x)+z^{\Delta}\frac{\langle O(x)\rangle}{2\Delta-4},
\end{equation}
where $\Phi_0(x)$ acts as a source for $O(x)$ and $\langle
O(x)\rangle$ denotes the corresponding condensate. It is seen that
at $\Delta=2$ the relation~\eqref{25a} is not well defined.

The fact that the no-wall model describes the discrete spectrum of
axial-vector mesons and not of the vector mesons can be demonstrated
straightforwardly. Following~\cite{son1,pom}, let us introduce the
left ($L$) and right ($R$) 5D vector fields corresponding on the 4D
boundary to the sources for the left and right vector currents,
$A_L^M(x,\epsilon)\leftrightarrow\bar{q}_L\gamma^{\mu}q_L$ and
$A_R^M(x,\epsilon)\leftrightarrow\bar{q}_R\gamma^{\mu}q_R$. They are
related to the usual $V$ and $A$ fields as $V=A_L+A_R$, $A=A_L-A_R$.
Then the action of no-wall model is
\begin{equation}
\label{28a}
S=\int d^4\!x\,dz\sqrt{g}\left(|D_M\varphi|^2+4\varphi^2
-\frac14F_L^2-\frac14F_R^2\right),
\end{equation}
Since the reflection of coordinate means in 5D space the interchange
of left and right fields, the spatial parity dictates the following
form for the covariant derivative~\cite{son1,pom}:
$D_M=\partial_M-i\lambda(A_{L,M}-A_{R,M})$. The action takes the form
\begin{equation}
\label{29a}
S=\int d^4\!x\,dz\sqrt{g}\left(|\partial_M\varphi-i\lambda
A_M\varphi|^2+4\varphi^2 -\frac18F_A^2-\frac18F_V^2\right).
\end{equation}
As follows from~\eqref{29a}, the spectrum of $A$-mesons will be discrete, $m_A^2\sim n+1$,
while the spectrum of $V$-mesons will be continuous.

Obviously, the case of $U_L(1)\times U_R(1)$ gauge symmetry
discussed above can be generalized to the $SU_L(2)\times SU_R(2)$
case.

\section{The scalar sector}

Let us consider the free scalar action of the SW model,
\begin{equation}
\label{31a}
S=\int d^4\!x\,dz\sqrt{g}\,e^{-az^2}(\partial_M\Phi\partial^M\Phi-m^2\Phi^2).
\end{equation}
The ensuing equation of motion results in the following spectrum,
\begin{equation}
\label{33a}
m_n^2=2|a|\left(2n+1+\sqrt{4+m_5^2}+\frac{a}{|a|}\right),\qquad n=0,1,2,\dots.
\end{equation}
We are interested in the situation when the massless scalar state
appears. There is only one possibility: $a<0$, $m_5^2=-4$. As we
indicated above, the value $m_5^2=-4$ corresponds to the minimal
value for the mass squared in the AdS space~\cite{freedman}. This
possibility has been mentioned in Ref.~\cite{br3}.

Following our proposal in Section~2, the choice $a<0$ should correspond
to the axial mesons in the vector channel and to the pseudoscalar
mesons in the scalar one. It looks encouraging to observe that with
this choice of sign the appearance of the massless pole in the
vector correlator may be accompanied by the appearance of the
massless state in the scalar part of the action. It is interesting
to remark also that the comparison of the ensuing pseudoscalar spectrum
\begin{equation}
\label{34a}
m_n^2=4|a|n,
\end{equation}
with the axial spectrum, $m_n^2=4|a|(n+1)$, leads to the prediction
that $m_{\pi'}=m_{a_1}$, where $\pi'$ denotes the first radial
excitation of the pion. This prediction is compatible with the
experiment~\cite{pdg}: $m_{\pi'}=1300\pm100$~MeV,
$m_{a_1}=1230\pm40$~MeV.

\section{Discussions}

As has been noticed in Ref.~\cite{son3}, aside from the existence of
massless state in the vector channel, there is an additional
argument against the $a<0$ choice in the SW models: it would give a
higher spin meson spectrum independent of spin $j$,
$m_{n,j}^2=4|a|(n+1)$, while the SW model with the positive sign
$a>0$ gives the expected string like spectrum
$m_{n,j}^2=4|a|(n+j)$~\cite{son2}. We wish to make a couple of
comments on this arguments. First, it does not exclude the proposal
of the present work since the negative sign $a<0$ can be viewed as a
peculiarity of the axial-vector channel. Second, this argument seems
to be valid for the gauge higher spin fields in AdS$_5$ which can be
made massless by a special choice of the gauge conditions. However,
such an introduction of higher spin fields is not unique. For
instance, Brodsky and Teramond advocated in the Ref.~\cite{br2} that
the relation~\eqref{3a} for the 5D mass extends to the $J>1$ mesons.
In this case, the argument above is not valid (see,
e.g., Ref.~\cite{gutsche}).

It should be emphasized that the massless pole in the axial-vector
correlator does not imply the existence of massless axial-vector
state. This can be seen as follows. The physical states correspond
to normalizable solutions of the kind~\eqref{13a}. The corresponding
eigenfunctions $v_n(z)$ form a complete set of functions. Thus, it
is possible to make the expansion $A^{\mu}(x,z)=\sum_n
A^{\mu}_n(x)v_n(z)$. Substituting this expansion back into the original
action and integrating over $z$ one arrives at a 4D action
containing the infinite number of free fields with masses squared
$m_n^2$ given by the eigenvalues corresponding to the eigenfunctions
$v_n(z)$ (the procedure is written in detail in Ref.~\cite{afoninI}),
\begin{equation}
\label{36a}
S=c^2\int d^4\!x\sum_{n=0}^{\infty}\left[-\frac14\left(F_{\mu\nu}^{(n)}\right)^2+m_n^2\left(A_{\mu}^{(n)}\right)^2\right]
\end{equation}
The massless state does not appear in the 4D action since it is
non-normalizable.

We note finally that the existence of a massless non-normalizable
solution that leaves the action finite, in the general case, does
not lead to the massless pole in the corresponding correlator. For
instance, the Eq.~\eqref{5a} has the second massless
non-normalizable solution, $v(z)=\text{const}$, which leaves the
action~\eqref{6a} finite at any $a$ (namely, giving zero
contribution to the action). This fact becomes especially evident if
one considers a 5D action for the scalar fields or non-gauge vector
fields\footnote{Because in contrast to the definition~\eqref{7a}, in
these cases one does not extract the factor $q^2$ in defining the
two-point correlator.}, where such solutions can yield a non-zero
contribution to the action.

\section{Conclusions}

The chiral symmetry breaking is usually introduced into the
bottom-up holographic models by hands via an extra scalar field or
special boundary conditions. This strategy in the search for
successful dynamical model is inherited from the low-energy
effective field theories for QCD. As long as the foundations of
holographic correspondence are not yet well understood, it is not
clear why we may separate the dynamics responsible for the formation
of masses from the corrections to this dynamics caused by the chiral
symmetry breaking. In fact, building the five-dimensional dual
models we cannot {\it a priori} control such a separation. We have
argued in the present work that in the axial-vector sector, the soft
wall model with the exponentially growing quadratic background
includes a substantial part of the CSB already on the level of free
fields\footnote{As a consequence, the standard introduction of the
CSB via adding the 5D scalar field dual to the quark operator
$\bar{q}q$ may meet the double counting problem.}. The massless pole
in the correlator of conserved vector currents (not corresponding to
a real massless vector particle) actually means that the given
current is conserved due to the presence of a massless scalar $\pi$
such that $A_{\mu}\sim\partial_{\mu}\pi$. Thus, the PCAC hypothesis
becomes a prediction of the model. In addition, the spectrum of the
model resembles much more the spectrum of known axial states than
the vector one.

Aside from the massless pole in the axial correlator, an interesting
unexpected property of changing the dilaton sign is the possibility
to have automatically the massless state in the scalar channel. In
summary: The standard soft-wall model seems to be good for the
description of the vector and scalar mesons while for the
axial-vector and pseudoscalar channels it looks better to change the
sign of the dilaton background.

\section*{Acknowledgments}

The work is supported by SPbSU grant 11.0.64.2010 and by the Dynasty
Foundation.

\end{document}